\documentstyle[12pt]{article}

\begin{document}

\begin{center}
{\bf
FLUCTUATIONS IN THE MIXED EVENT TECHNIQUE
}\\
\bigskip
\bigskip
Sergei VOLOSHIN\footnote{ On leave from Moscow Engineering Physics Institute,
     Moscow, 115409,  Russia}

\bigskip
{\em
University of Pittsburgh, Pittsburgh, PA 15260
}\\

\end{center}

\bigskip
{\footnotesize
\centerline{ABSTRACT}
\begin{quotation}
\vspace{-0.10in}
%
A method for an evaluation of fluctuations in the mixed event
technique is proposed. It is shown, that, generally, the magnitude
of the fluctuations is proportional to $N^{3/4}$, where $N$ is the number of
produced events, which should be compared with $N^{1/2}$ for the case
of independent events. The formula eligible for the use in an analysis
of experimental data is provided.
\vspace{7pt}
\newline
PACS number(s): 13.85.-t, 13.85.Hd, 25.70.Pq
\end{quotation}}
\bigskip

The mixed event technique was initially proposed [1] for the generation of
uncorrelated pion pairs for the intensity interferometry of identical bosons.
Later on, the technique was widely used for the generation of background
distributions for different normalized correlation functions and in many
other applications, {\it e.g.} for the calculation of
background mass spectrum in resonance reconstruction.
To use all the available statistic very often the same particle is used
in generation of many ``events'';
 thus produced ``events'' are not totally independent.
Our goal is to evaluate the fluctuations in the produced distributions.
In this note we consider particularly the case of the distribution of pairs
of particles in their relative momentum, but the method can be applied to
any distribution generated by event mixing.

Below we derive the formula for the variance of statistical fluctuation
in the number of pairs in some region of the particles relative momentum.
To generate pairs we use $M$ pions from {\em $M$ different} events
(one pion from each event), generating altogether
$N_{pairs}^{tot}=M(M-1)/2$
pairs, distributed among different bins in
accordance with the value of relative momentum
$q_{ij}=p_i-p_j;\;i,j=1,2,...,M$.
We study the fluctuations in one particular bin.
It is convenient to introduce the function
\begin{equation}
n(p_i-p_j) \equiv n_{ij}=\left\{ \begin{array}{ll}
         1   & \mbox{if $q_{ij}$ belongs to the bin} \\
         0   & \mbox{otherwise}
         \end{array}
  \right.
\label{enij}
\end{equation}

Using this notation the mean number of pairs in the bin under study
\begin{equation}
\overline{N_{pairs}^{bin}}=N_{pairs}^{tot}\overline{n_{ij}}
\equiv N_{pairs}^{tot}\overline{n} .
\end{equation}
Note that $\overline{n}$ depends only on the particle distribution
in momentum and bin size
and position, but does not depend on $M$!

$N_{pairs}^{bin}$ is a random variable
(below we use the shorter notation $X \equiv N_{pairs}^{bin}$)
with a mean value $\overline{X}$  and variance $\sigma_X$;
\begin{equation}
X  =\frac{1}{2} \sum_{i \neq j} n_{ij};
\end{equation}
To get the variance we have to calculate
\begin{eqnarray}
\sigma_X^2 &=& \overline{X^2}-\overline{X}^2=  \nonumber \\
          &=& \overline{(\frac{1}{2} \sum_{i \neq j} n_{ij})^2}
                -(\frac{M(M-1)}{2})^2 \overline{n}^2=   \nonumber \\
       &=& \frac{1}{4} \{ M(M-1)(M-2)(M-3)
        (\overline{n_{ij}n_{kl}} -\overline{n}^2)+
                                                          \nonumber \\
       & & +4M(M-1)(M-2)(\overline{n_{ik}n_{kl}} -\overline{n}^2)+
                                                          \nonumber \\
       & & +2M(M-1)(\overline{n_{ik}^2} -\overline{n}^2) \}.
\end{eqnarray}

In the first term of the last expression all four
indices ($i,j,k,l$) are different; consequently
$\overline{n_{ij}n_{kl}}=\overline{n}^2$.
Due to this  the first term equals to zero and does not contribute to
the variance.
Let us go now to the last term. Its contribution is
\begin{equation}
\frac{2M(M-1)}{4} [\overline{n_{ik}^2} -\overline{n}^2]=
\overline{X}(1-\overline{n}),
\end{equation}
where we used the equality $\overline{n_{ik}^2}=\overline{n}$,
following from the definition (\ref{enij}).
Usually $\overline{n} \ll 1$, thus the contribution of the last term to
the variance is just the mean number of pairs in the bin; the value which
we would expect if the pairs were totally independent.
But it is known that the pairs are really
correlated because each pion is used in generation of $(M-1)$ pairs.
The price for this is the second term in the expression for the variance.
Let us denote by $a=\overline{n_{ik}n_{kl}} -\overline{n}^2$.
Note that this value, like $\overline{n}$, does not depend on $M$.
Then for the variance we have:
\begin{equation}
\sigma_X^2 =  M(M-1)(M-2) a - \overline{X}(1-\overline{n_{ij}})
   \approx M^3 a-  \overline{X}.
\end{equation}
If $aM\gg  \overline{n}$ the first term becomes dominant and
\begin{equation}
\sigma_X^2 \approx a (2/ \overline{n})^{3/2} \overline{X}^{3/2}
\end{equation}
The fluctuations in this case are proportional to $(N_{pairs}^{bin})^{3/4}$
, and can be much larger than for the case of totally independent
pairs\footnote{This is
exactly the law seen by Zajc et al. [2] {\em empirically}.
In the Ref.[2] this law was derived in
a simple model, but for the real case the authors
studied empirical relations.}.

The parameters $a$ and $\overline{n}$ can be expressed using
one-particle density:
\begin{equation}
\overline{n}=\frac{1}{\overline{N} ^2}\int dp_i dp_j \rho^{(1)}(p_i)
      \rho^{(1)}(p_j)n_{ij};
\end{equation}
\begin{equation}
a=\overline{n}^2+\frac{1}{\overline{N} ^3}
     \int dp_i dp_l dp_k \rho^{(1)}(p_i)
                         \rho^{(1)}(p_l)
                         \rho^{(1)}(p_k)n_{ik}n_{kl},
\end{equation}
where $\overline{N}$ is the mean number of particles (the value one-particle
density is normalized to).
We used the fact that the individual particles in our technique
are independent, and we can express two- and three-particle
densities as a product of one-particle densities.

In the analysis of the experimental data these parameters should be substituted
by their estimators
(denoted by $\langle \rangle$)
directly using the data. For example, one can use
\begin{equation}
\overline{n}\approx \langle n_{ij} \rangle =
\frac{1}{M(M-1)} \sum_{i \neq j} n_{ij}
\end{equation}
\begin{equation}
a \approx \langle n_{ik}  n_{kl}\rangle-\langle n_{ij} \rangle^2
= \frac{1}{M(M-1)(M-2)} \sum_{i,k,l}  n_{ik}n_{kl}-\langle n_{ij} \rangle^2,
\end{equation}
where in the last sum all three indices ($i,k,l$) are different.
Since $a$ does not depend on $M$, the average does not need to be carried
out over the entire sample.

\medskip
Once we have seen that the fluctuations in the number of mixed pairs
can be rather large it becomes
worthwhile to recall that the estimator usually used for the correlation
function defined as a ratio of estimators for the numbers of the true and mixed
pairs is biased:
\begin{equation}
\overline{\langle C_2 \rangle}=
\overline{\frac{N_{pairs}^{true} }{ N_{pairs}^{mixed} }}
\neq
C_2^{true}=
\frac{ \overline{N_{pairs}^{true}} }{\overline{ N_{pairs}^{mixed}} }.
\end{equation}
The necessary correction can be derived from the equality
\begin{equation}
\overline{\frac{1}{X}}=\overline{\frac{1}{\overline{X}+(X-\overline{X})}} =
\frac{1}{\overline{X}}(1+\frac{\overline{(X-\overline{X})^2}}{\overline{X}^2}-
\frac{\overline{(X-\overline{X})^3}}{\overline{X}^3}+...)
\approx
\frac{1}{\overline{X}}(1+\frac{\sigma_X^2}{\overline{X}^2}),
\end{equation}
where $\sigma_X^2$ is defined by Eq.6. The corrected estimator for
the correlation function is
\begin{equation}
\langle C_2 \rangle^{corrected}=\langle C_2 \rangle
(1-\frac{\sigma_X^2}{X^2}),
\end{equation}
where, as earlier, we denote by $X$ the mean value of mixed pairs in the
bin under study.

\bigskip

The author thanks Nu Xu for drawing his attention to the problem
and Wilfred Cleland for fruitful discussions.
This work was supported by the US DoE under Contract No. DEFG02-87ER40363.
\bigskip
\bigskip

{\bf Reference}
\bigskip

[1] G. I. Kopylov, Phys. Lett. {\bf 50B}, 472 (1974)

[2] W. A. Zajc et al., Phys. Rev. {\bf C 29}, 2173 (1984)

\end{document}